\documentstyle[preprint,psfig,floats,pra,aps]{revtex}
\tightenlines
\begin{document} \draft

\title{\Large \bf Two-pearl Strings: Feynman's Oscillators}

\author{Y. S. Kim\footnote{electronic mail: yskim@physics.umd.edu}}
\address{Department of Physics, University of Maryland, College Park,
Maryland 20742}

\maketitle

\begin{abstract}

String models are designed to provide a covariant description of
internal space-time structure of relativistic particles.  The string
is a limiting case of a series of massive beads like a pearl necklace.
In the limit of infinite-number of zero-mass beads, it becomes a
field-theoretic string.  Another interesting limit is to keep only
two pearls by eliminating all others, resulting in a harmonic
oscillator.  The basic strength of the oscillator model is its
mathematical simplicity.  This encourages us to construct two-pearl
strings for a covariant picture of relativistic extended particles.
We achieve this goal by transforming the oscillator model of
Feynman {\it et al.} into a representation of the Poincar\'e group.
We then construct representations of the $O(3)$-like little group for
those oscillator states, which dictates their internal space-time
symmetry of massive particles.  This simple mathematical procedure
allows us to explain what we observe in the world in terms of the
fundamental space-time symmetries, and the built-in covariance of
the model allows us to use the physics in the rest frame in order
to explain what happens in the infinite-momentum frame.  It is thus
possible to calculate the parton distribution within the proton
moving light-like speed in terms of the quark wave function in its
rest frame.

\end{abstract}

\pacs{  }

\section{Introduction}\label{intro}
Physicists are fond of building strings.  In classical mechanics, we
start with a discrete set of particles joined together with a finite
distance between two neighboring particles, like a pearl necklace.
We then take the limit of zero distance and infinite number of
particles, resulting in a continuous string.  This is how we construct
classical field theory and then extend it to quantum field theory in
the Lagrangian formalism.  In this paper, we consider the opposite
limit by dropping all the particles except two.

In order to gain an insight into what we intend to in this report, let
us note an example in history.  Debye's treatment of specific heat is
a classic example.  Einstein's oscillator model of specific heat is a
simplified case of the Debye model in the sense that it consists only
of two pearls.  The Einstein model does not give an accurate
description of the specific heat in the zero-temperature limit, but
it is accurate enough everywhere else to be covered in textbooks.  The
basic strength of the oscillator model is its mathematical simplicity.
It produces the numbers and curves which can be checked experimentally,
without requiring from us too much mathematical labor.

While one of the main purposes of the string models is to study the
internal space-time symmetries of relativistic particles, we can
achieve this purpose by studying two-pearl strings which should share
the same symmetry property as all other string models.  In practice,
the two-pearl string model consists of two constituents joined together
by a spring force.  The only problem is to construct the oscillator
model which can be Lorentz-transformed.  The problem then is to reduced
to constructing a covariant harmonic oscillator formalism.  This
subject has a long history~\cite{dir45,yuka53,markov56,fkr71}.

In Ref.~\cite{fkr71}, Feynman {\it et al.} attempted to construct a
covariant model for hadrons consisting of quarks joined together by
an oscillator force.  They indeed formulated a Lorentz-invariant
oscillator equation.  They also worked out the degeneracies of the
oscillator states which are consistent with observed mesonic and
baryonic mass spectra.  However, their wave functions are not
normalizable in the space-time coordinate system.  The authors of
this paper never considered the question of covariance.

What is the relevant question on covariance within the framework of
the oscillator formalism?  In 1969~\cite{fey69}, Feynman proposed his
parton model for hadrons moving with almost speed of light.  Feynman
observed that the hadron consists of collection of infinite number of
partons which are like free particles.  The partons appear to have
properties which are quite different from those of the quarks.  If the
wave functions are to be covariant, they should be able to translate
the quark model for slow hadrons into the parton model of fast hadrons.
This is precisely the question we would like to address in the present
report.

We achieve this purpose by transforming the oscillator model of
Feynman {\it et al.} into a representation of the Poincar\'e group
which governs the space-time symmetries of relativistic
particles~\cite{wig39,knp86}.  In this formalism, the internal
space-time symmetries are dictated by the little groups.  The
little group is the maximal subgroup of the Lorentz group whose
transformations leave the four-momentum of a given particle
invariant.  The little groups for massive and massless particles
are known to be isomorphic to $O(3)$ or the three-dimensional
rotation group and $E(2)$ or the two-dimensional Euclidean
group~\cite{wig39,knp86}.  In this paper, we can rewrite the wave
functions of Feynman {\it et al.} as a representation of the
$O(3)$-like little group for a massive.

Let us go back to physics.  When Einstein formulated $E = mc^{2}$ in
1905, he was talking about point particles.  These days, particles
have their own internal space-time structures.  In the case of hadrons,
the particle has a space-time extension like the hydrogen atom.  In
spite of these complications, we do not question the validity of the
energy-momentum relation given by $E = \sqrt{m^{2} + p^{2}} $ for all
relativistic particles.  The problem is that each particle has its own
internal space-time variables.  In addition to the energy and momentum,
the massive particle has a package of variables including mass, spin,
and quark degrees of freedom.  The massless particle has its helicity,
gauge degrees of freedom, and parton degrees of freedom.

The question is whether the two different packages of new variables
for massive and massless particles can be combined into a single
covariant package as Einstein's $E = mc^{2}$ does for the
energy-momentum relations for massive and massless particles.  We
shall divide this question into two parts.  First, we deal with
the question of spin, helicity, and gauge degrees of freedom.  We
can deal with this question without worrying about the space-time
extension of the particle.  Second, we face the problem of
space-time extensions using hadrons which are bound states of
quarks obeying the laws of quantum mechanics.  In order to answer
this question, we first have to construct a quantum mechanics of
bound states which can be Lorentz-boosted.

In Sec.~\ref{problem}, the above-mentioned problems are spelled out
in detail.
In Sec.~\ref{littleg}, we present a brief history of applications of
the little groups of the Poincar\'e to internal space-time symmetries
of relativistic particles.  In Sec.~\ref{covham}, we construct
representations of the little group using harmonic oscillator wave
functions.
In Sec.~\ref{parton}, it is shown that the Lorentz-boosted oscillator
wave functions exhibit the peculiarities Feynman's parton model in
the infinite-momentum limit.

Much of the concept of Lorentz-squeezed wave function is derived from
elliptic deformations of a sphere resulting in a mathematical technique
group called contractions~\cite{inonu53}.  In Appendix~\ref{o3e2}, we
discuss the contraction of the three-dimensional rotation group to the
two-dimensional Euclidean group.  In Appendix~\ref{contrac}, we
discuss the little group for a massless particle as the
infinite-momentum/zero-mass limit of the little group for a massive
particle.

\section{Statement of the Problem}\label{problem}

The Lorentz-invariant differential equation of Feynman, Kislinger, and
Ravndal is a linear partial differential equation~\cite{fkr71} .  It
can therefore generate many different sets of solutions depending on
boundary conditions.  In their paper, Feynman {\it et al.} choose
Lorentz-invariant solutions.  But their solutions are not normalizable
and cannot therefore be interpreted within the framework of the existing
rules of quantum mechanics.  In this report, we point out there are
other sets of solutions.  We choose here normalizable wave functions.
They are not Lorentz-invariant, but they are Lorentz-covariant.  These
covariant solutions form a representations of the Poincar\'e
group~\cite{wig39,knp86}.

The Lorentz-invariant wave function takes the same form in every
Lorentz frame, but the covariant wave function takes different forms.
However, in the covariant formulation, the wave function in one frame
can be transformed to the wave function in a different frame by
Lorentz transformation.  In particular, the wave function in the
infinite-momentum frame is quite different from the wave function at
the rest frame.  Thus, it may be possible to obtain Feynman's parton
picture by Lorentz-boosting the quark wave function constructed from
the rest frame.

In spite of the mathematical difficulties, the original paper of
Feynman {\it et al.} contains the following radical departures from
the conventional viewpoint.

\begin{itemize}

\item For relativistic bound state, we should use harmonic oscillators
instead of Feynman diagrams.

\item  We should us harmonic oscillators instead of Regge trajectories
to study degeneracies in the hadronic spectra.

\end{itemize}

These views sound radical, but they are quite consistent with the
existing forms of quantum mechanics and quantum field theory.  In
quantum field theory, Feynman diagrams are only for scattering states
where the external lines correspond free particles in asymptotic
states.  The oscillator eigenvalues are proportional to the highest
values of the angular momentum.  This is often known as the linear
Regge trajectory.  Between the Regge trajectory and the
three-dimensional oscillator, which one is closer to the fundamental
laws of quantum mechanics.  Therefore, the above-mentioned radical
departures mean that we are coming back to common sense in physics.

On the other hand, there is one important point Feynman {\it et al.}
failed to see in their oscillator paper~\cite{fkr71}.  Two years
before the publication of this oscillator paper, Feynman proposed his
parton model~\cite{fey69}.  However, in their oscillator paper, they
do not mention the possibility of obtaining the parton picture from
the quantum mechanics of bound-state quarks in a hadron in its rest
frame.  It is probably because their wave functions are
Lorentz-invariant but not covariant.

However, the covariant formalism forces us to raise this question.
This is precisely the purpose of the present report.

\section{Poincar\'e Symmetry of Relativistic Particles}\label{littleg}

The Poincar\'e group is the group of inhomogeneous Lorentz
transformations, namely Lorentz transformations preceded or followed
by space-time translations.  In order to study this group, we have to
understand first the group of Lorentz transformations, the group of
translations, and how these two groups are combined to form the
Poincar\'e group.  The Poincar\'e group is a semi-direct product of
the Lorentz and translation groups.  The two Casimir operators of
this group correspond to the (mass)$^{2}$ and (spin)$^{2}$ of a given
particle.  Indeed, the particle mass and its spin magnitude are
Lorentz-invariant quantities.

The question then is how to construct the representations of the
Lorentz group which are relevant to physics.  For this purpose,
Wigner in 1939 studied the subgroups of the Lorentz group whose
transformations leave the four-momentum of a given free
particle~\cite{wig39}.  The maximal subgroup of the Lorentz group
which leaves the four-momentum invariant is called the little group.
Since the little group leaves the four-momentum invariant, it governs
the internal space-time symmetries of relativistic particles.  Wigner
shows in his paper that the internal space-time symmetries of massive
and massless particles are dictated by the $O(3)$-like and $E(2)$-like
little groups respectively.

The $O(3)$-like little group is locally isomorphic to the
three-dimensional rotation group, which is very familiar to us.
For instance, the group $SU(2)$ for the electron spin is an $O(3)$-like
little group.  The group $E(2)$ is the Euclidean group in a
two-dimensional space, consisting of translations and rotations on a
flat surface.  We are performing these transformations everyday on
ourselves when we move from home to school.  The mathematics of these
Euclidean transformations are also simple.  However, the group of
these transformations are not well known to us.  In Appendix \ref{o3e2},
we give a matrix representation of the $E(2)$ group.

The group of Lorentz transformations consists of three boosts and
three rotations.  The rotations therefore constitute a subgroup of
the Lorentz group.  If a massive particle is at rest, its four-momentum
is invariant under rotations.  Thus the little group for a massive
particle at rest is the three-dimensional rotation group.  Then what is
affected by the rotation?  The answer to this question is very simple.
The particle in general has its spin.  The spin orientation is going
to be affected by the rotation!

If the rest-particle is boosted along the $z$ direction, it will pick
up a non-zero momentum component.  The generators of the $O(3)$ group
will then be boosted.  The boost will take the form of conjugation by
the boost operator.  This boost will not change the Lie algebra of the
rotation group, and the boosted little group will still leave the
boosted four-momentum invariant.  We call this the $O(3)$-like little
group.  If we use the four-vector coordinate $(x, y, z, t)$, the
four-momentum vector for the particle at rest is $(0, 0, 0, m)$, and
the three-dimensional rotation group leaves this four-momentum
invariant.  This little group is generated by
\begin{equation}
J_{1} = \pmatrix{0&0&0&0\cr0&0&-i&0\cr0&i&0&0\cr0&0&0&0} , \qquad
J_{2} = \pmatrix{0&0&i&0\cr0&0&0&0\cr-i&0&0&0\cr0&0&0&0} ,
\end{equation}
and
\begin{equation}\label{j3}
J_{3} = \pmatrix{0 & -i & 0 & 0 \cr i & 0 & 0 & 0
\cr 0 & 0 & 0 & 0 \cr 0 & 0 & 0 & 0} ,
\end{equation}
which satisfy the commutation relations:
\begin{equation}
[J_{i}, J_{j}] = i\epsilon_{ijk} J_{k} .
\end{equation}

It is not possible to bring a massless particle to its rest frame.
In his 1939 paper~\cite{wig39}, Wigner observed that the little group
for a massless particle moving along the $z$ axis is generated by the
rotation generator around the $z$ axis, namely $J_{3}$ of Eq.(\ref{j3}),
and two other generators which take the form
\begin{equation}\label{n1n2}
N_{1} = \pmatrix{0 & 0 & -i & i \cr 0 & 0 & 0 & 0
\cr i & 0 & 0 & 0 \cr i & 0 & 0 & 0} ,  \quad
N_{2} = \pmatrix{0 & 0 & 0 & 0 \cr 0 & 0 & -i & i
\cr 0 & i & 0 & 0 \cr 0 & i & 0 & 0} .
\end{equation}
If we use $K_{i}$ for the boost generator along the i-th axis, these
matrices can be written as
\begin{equation}
N_{1} = K_{1} - J_{2} , \qquad N_{2} = K_{2} + J_{1} ,
\end{equation}
with
\begin{equation}
K_{1} = \pmatrix{0&0&0&i\cr0&0&0&0\cr0&0&0&0\cr i&0&0&0} , \qquad
K_{2} = \pmatrix{0&0&0&0\cr0&0&0&i\cr0&0&0&0\cr0&i&0&0} .
\end{equation}
The generators $J_{3}, N_{1}$ and $N_{2}$ satisfy the following set
of commutation relations.
\begin{equation}\label{e2lcom}
[N_{1}, N_{2}] = 0 , \quad [J_{3}, N_{1}] = iN_{2} ,
\quad [J_{3}, N_{2}] = -iN_{1} .
\end{equation}
In Appendix \ref{o3e2}, we discuss the generators of the $E(2)$ group.
They are $J_{3}$ which generates rotations around the $z$ axis, and
$P_{1}$ and $P_{2}$ which generate translations along the $x$ and $y$
directions respectively.  If we replace $N_{1}$ and $N_{2}$ by $P_{1}$
and $P_{2}$, the above set of commutation relations becomes the set
given for the $E(2)$ group given in Eq.(\ref{e2com}).  This is the
reason why we say the little group for massless particles is
$E(2)$-like.  Very clearly, the matrices $N_{1}$ and $N_{2}$ generate
Lorentz transformations.

It is not difficult to associate the rotation generator $J_{3}$ with
the helicity degree of freedom of the massless particle.   Then what
physical variable is associated with the $N_{1}$ and $N_{2}$
generators?  Indeed, Wigner was the one who discovered the existence
of these generators, but did not give any physical interpretation to
these translation-like generators.  For this reason, for many years,
only those representations with the zero-eigenvalues of the $N$
operators were thought to be physically meaningful
representations~\cite{wein64}.  It was not until 1971 when Janner
and Janssen reported that the transformations generated by these
operators are gauge transformations~\cite{janner71,kim97poz}.  The
role of this translation-like transformation has also been studied
for spin-1/2 particles, and it was concluded that the polarization
of neutrinos is due to gauge invariance~\cite{hks82,kim97min}.

\begin{table}

\caption{Further contents of Einstein's $E = mc^{2}$.  Massive and
massless particles have different energy-momentum relations.  Einstein's
special relativity gives one relation for both.  Wigner's little group
unifies the internal space-time symmetries for massive and massless
particles which are locally isomorphic to $O(3)$ and $E(2)$ respectively.
It is a great challenge for us to find another unification.  In this
note, we present a unified picture of the quark and parton models which
are applicable to slow and ultra-fast hadrons respectively.}

\vspace{3mm}

\begin{tabular}{cccc}
{}&{}&{}&{}\\
{} & Massive, Slow \hspace*{1mm} & COVARIANCE \hspace*{1mm}&
Massless, Fast \\[4mm]\hline
{}&{}&{}&{}\\
Energy- & {}  & Einstein's & {} \\
Momentum & $E = p^{2}/2m$ & $ E = [p^{2} + m^{2}]^{1/2}$ & $E = cp$
\\[4mm]\hline
{}&{}&{}&{}\\
Internal & $S_{3}$ & {}  &  $S_{3}$ \\[-1mm]
space-time &{} & Wigner's  & {} \\ [-1mm]
symmetry & $S_{1}, S_{2}$ & Little Group & Gauge
Transformations \\[4mm]\hline
{}&{}&{}&{}\\
Relativistic & {} & {} & {} \\[-1mm]
Extended & Quark Model & Covariant Model of Hadrons & Partons \\ [-1mm]
Particles & {} & {} & {} \\[2mm]
\end{tabular}
\end{table}

Another important development along this line of research is the
application of group contractions to the unifications of the two
different little groups for massive and massless particles.  We
always associate the three-dimensional rotation group with a spherical
surface.  Let us consider a circular area of radius 1 kilometer centered
on the north pole of the earth.  Since the radius of the earth is more
than 6,450 times longer, the circular region appears flat.  Thus, within
this region, we use the $E(2)$ symmetry group for this region.  The
validity of this approximation depends on the ratio of the two radii.

In 1953, Inonu and Wigner formulated this problem as the contraction of
$O(3)$ to $E(2)$~\cite{inonu53}.  How about then the little groups which
are isomorphic to $O(3)$ and $E(2)$?  It is reasonable to expect that
the $E(2)$-like little group be obtained as a limiting case for of the
$O(3)$-like little group for massless particles.  In 1981, it was
observed by Ferrara and Savoy that this limiting process is the Lorentz
boost~\cite{ferrara82}.  In 1983, using the
same limiting process as that of Ferrara and Savoy, Han {\it et al}
showed that transverse rotation generators become the generators of
gauge transformations in the limit of infinite momentum and/or zero
mass~\cite{hks83pl}.  In 1987, Kim and Wigner showed that the little
group for
massless particles is the cylindrical group which is isomorphic to the
$E(2)$ group~\cite{kiwi87jm}.  This completes the second raw in Table I,
where Wigner's little group unifies the internal space-time symmetries
of massive and massless particles.

We are now interested in constructing the third row in Table I.  As we
promised in Sec.~\ref{intro}, we will be dealing with hadrons which
are bound states of quarks with space-time extensions.  For this
purpose, we need a set of covariant wave functions consistent with the
existing laws of quantum mechanics, including of course the uncertainty
principle and probability interpretation.

With these wave functions, we propose to solve the following problem
in high-energy physics.  The quark model works well when hadrons are
at rest or move slowly.  However, when they move with speed close to
that of light, they appear as a collection of infinite-number of
partons~\cite{fey69}.  As we stated above, we need a set of wave
functions which can be Lorentz-boosted.  How can we then construct
such a set?  In constructing wave functions for any purpose in quantum
mechanics, the standard procedure is to try first harmonic oscillator
wave functions.  In studying the Lorentz boost, the standard language
is the Lorentz group.  Thus the first step to construct covariant wave
functions is to work out representations of the Lorentz group using
harmonic oscillators~\cite{dir45,yuka53,knp86}.

\section{Covariant Harmonic Oscillators}\label{covham}

If we construct a representation of the Lorentz group using normalizable
harmonic oscillator wave functions, the result is the covariant harmonic
oscillator formalism~\cite{knp86}.  The formalism constitutes a
representation of Wigner's $O(3)$-like little group for a massive
particle with internal space-time structure.  This oscillator formalism
has been shown to be effective in explaining the basic phenomenological
features of relativistic extended hadrons observed in high-energy
laboratories.  In particular, the formalism shows that the quark model
and Feynman's parton picture are two different manifestations of one
covariant entity~\cite{knp86,kim89}.  The essential feature of the
covariant harmonic oscillator formalism is that Lorentz boosts are
squeeze transformations~\cite{kn73,knp91}.  In the light-cone coordinate
system, the boost transformation expands one coordinate while contracting
the other so as to preserve the product of these two coordinate remains
constant.  We shall show that the parton picture emerges from this
squeeze effect.

Let us consider a bound state of two particles.  For convenience, we
shall call the bound state the hadron, and call its constituents quarks.
Then there is a Bohr-like radius measuring the space-like separation
between the quarks.  There is also a time-like separation between the
quarks, and this variable becomes mixed with the longitudinal spatial
separation as the hadron moves with a relativistic speed.  There are
no quantum excitations along the time-like direction.  On the other
hand, there is the time-energy uncertainty relation which allows
quantum transitions.  It is possible to accommodate these aspect within
the framework of the present form of quantum mechanics.  The uncertainty
relation between the time and energy variables is the c-number
relation~\cite{dir27},
which does not allow excitations along the time-like coordinate.  We
shall see that the covariant harmonic oscillator formalism accommodates
this narrow window in the present form of quantum mechanics.

For a hadron consisting of two quarks, we can consider their space-time
positions $x_{a}$ and $x_{b}$, and use the variables
\begin{equation}
X = (x_{a} + x_{b})/2 , \qquad x = (x_{a} - x_{b})/2\sqrt{2} .
\end{equation}
The four-vector $X$ specifies where the hadron is located in space and
time, while the variable $x$ measures the space-time separation between
the quarks.  In the convention of Feynman {\it et al.} \cite{fkr71},
the internal motion of the quarks bound by a harmonic oscillator
potential of unit strength can be described by the Lorentz-invariant
equation
\begin{equation}\label{osceq}
{1\over 2}\left\{x^{2}_{\mu} -
{\partial ^{2} \over \partial x_{\mu }^{2}}
\right\} \psi (x)= \lambda \psi (x) .
\end{equation}
It is now possible to construct a representation of the Poincar\'e group
from the solutions of the above differential equation~\cite{knp86}.

The coordinate $X$ is associated with the overall hadronic
four-momentum, and the space-time separation variable $x$ dictates
the internal space-time symmetry or the $O(3)$-like little group.  Thus,
we should construct the representation of the little group from the
solutions of the differential equation in Eq.(\ref{osceq}).  If the
hadron is at rest, we can separate the $t$ variable from the equation.
For this variable we can assign the ground-state wave function to
accommodate the c-number time-energy uncertainty relation~\cite{dir27}.
For the three space-like variables, we can solve the oscillator
equation in the spherical coordinate system with usual orbital and
radial excitations.  This will indeed constitute a representation of
the $O(3)$-like little group for each value of the mass.  The solution
should take the form
\begin{equation}
\psi (x,y,z,t) = \psi (x,y,z) \left({1\over \pi }\right)^{1/4}
\exp \left(-t^{2}/2 \right) ,
\end{equation}
where $\psi(x,y,z)$ is the wave function for the three-dimensional
oscillator with appropriate angular momentum quantum numbers.  Indeed,
the above wave function constitutes a representation of Wigner's
$O(3)$-like little group for a massive particle \cite{knp86}.

\begin{figure}[thb] 
\centerline{\psfig{figure=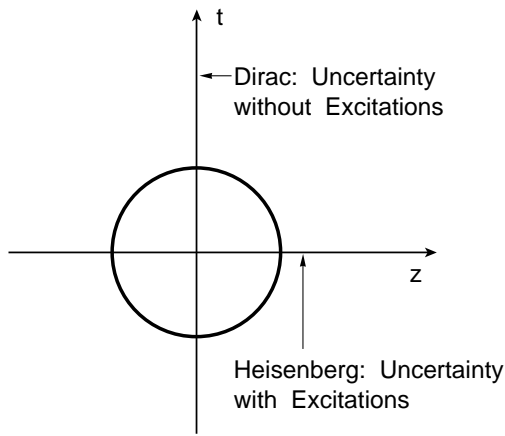,angle=0,height=60mm}}
\vspace{5mm}
\caption{Present form of quantum mechanics.  There are excitations
along the space-like dimensions, but there are no excitations along
the time-like direction.  However, there still is a time-energy
uncertainty relation.  We call this Dirac's c-number time-energy
uncertainty relation.  It is very important to note that this
space-time asymmetry is quite consistent with the concept of
covariance}\label{f.quantum}
\end{figure}

Since the three-dimensional oscillator differential equation is
separable in both spherical and Cartesian coordinate systems,
$\psi(x,y,z)$ consists of Hermite polynomials of $x, y$, and $z$.
If the Lorentz boost is made along the $z$ direction, the $x$ and $y$
coordinates are not affected, and can be temporarily dropped from the wave
function.  The wave function of interest can be written as
\begin{equation}
\psi^{n}(z,t) = \left({1\over \pi }\right)^{1/4}\exp \pmatrix{-t^{2}/2}
\psi_{n}(z) ,
\end{equation}
with
\begin{equation}
\psi ^{n}(z) = \left({1 \over \pi n!2^{n}} \right)^{1/2} H_{n}(z)
\exp (-z^{2}/2) ,
\end{equation}
where $\psi ^{n}(z)$ is for the $n$-th excited oscillator state.
The full wave function $\psi ^{n}(z,t)$ is
\begin{equation}\label{2.6}
\psi ^{n}_{0}(z,t) = \left({1\over \pi n! 2^{n}}\right)^{1/2} H_{n}(z)
\exp \left\{-{1\over 2}\left(z^{2} + t^{2} \right) \right\} .
\end{equation}
The subscript $0$ means that the wave function is for the hadron at
rest.  The above expression is not Lorentz-invariant, and its
localization undergoes a Lorentz squeeze as the hadron moves along the
$z$ direction~\cite{knp86}.  The above form of the wave function is
illustrated in Fig.\ref{f.quantum}.

\begin{figure}[thb] 
\centerline{\psfig{figure=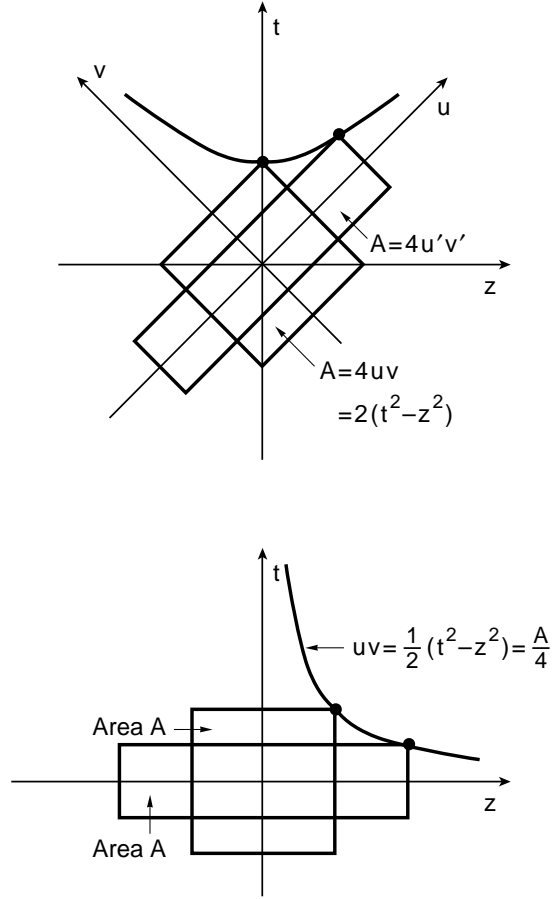,angle=0,height=120mm}}
\vspace{5mm}
\caption{Further contents of Lorentz boosts.  In the light-cone
coordinate system, the Lorentz boost takes the form of the lower part
of this figure.  In terms of the longitudinal and time-like variables,
the transformation is illustrated in the upper portion of this
figure.}\label{f.lcone}
\end{figure}

It is convenient to use the light-cone variables to describe Lorentz
boosts.  The light-cone coordinate variables are
\begin{equation}
u = (z + t)/\sqrt{2} , \qquad v = (z - t)/\sqrt{2} .
\end{equation}
In terms of these variables, the Lorentz boost along the $z$
direction,
\begin{equation}
\pmatrix{z' \cr t'} = \pmatrix{\cosh \eta & \sinh \eta \cr
\sinh \eta & \cosh \eta } \pmatrix{z \cr t} ,
\end{equation}
takes the simple form
\begin{equation}\label{lorensq}
u' = e^{\eta } u , \qquad v' = e^{-\eta } v ,
\end{equation}
where $\eta $ is the boost parameter and is $\tanh ^{-1}(v/c)$.
Indeed, the $u$ variable becomes expanded while the $v$ variable becomes
contracted.  This is the squeeze mechanism illustrated discussed
extensively in the literature~\cite{kn73,knp91}.  This squeeze
transformation is also illustrated in Fig.~\ref{f.lcone}.

The wave function of Eq.(\ref{2.6}) can be written as
\begin{equation}\label{10}
\psi ^{n}_{o}(z,t) = \psi ^{n}_{0}(z,t)
= \left({1 \over \pi n!2^{n}} \right)^{1/2} H_{n}\left((u + v)/\sqrt{2}
\right) \exp \left\{-{1\over 2} (u^{2} + v^{2}) \right\} .
\end{equation}
If the system is boosted, the wave function becomes
\begin{equation}\label{11}
\psi ^{n}_{\eta }(z,t) = \left({1 \over \pi n!2^{n}} \right)^{1/2}
H_{n} \left((e^{-\eta }u + e^{\eta }v)/\sqrt{2} \right)
\times \exp \left\{-{1\over 2}\left(e^{-2\eta }u^{2} +
e^{2\eta }v^{2}\right)\right\} .
\end{equation}


\begin{figure}[thb] 
\centerline{\psfig{figure=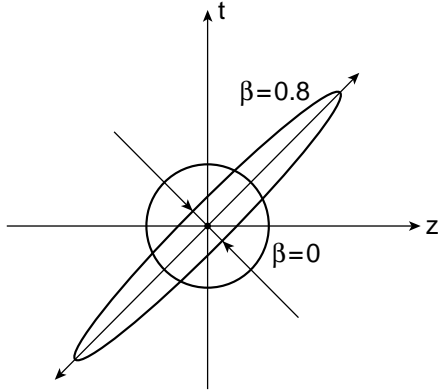,angle=0,height=60mm}}
\caption{Effect of the Lorentz boost on the space-time wave function.
The circular space-time distribution at the rest frame becomes
Lorentz-squeezed to become an elliptic distribution.}\label{f.ellipse}
\end{figure}


In both Eqs. (\ref{10}) and (\ref{11}), the localization property of
the wave function in the $u v$ plane is determined by the Gaussian factor,
and it is sufficient to study the ground state only for the essential
feature of the boundary condition.  The wave functions in Eq.(\ref{10})
and Eq.(\ref{11}) then respectively become
\begin{equation}\label{13}
\psi _{0}(z,t) = \left({1 \over \pi} \right)^{1/2}
\exp \left\{-{1\over 2} (u^{2} + v^{2}) \right\} .
\end{equation}
If the system is boosted, the wave function becomes
\begin{equation}\label{14}
\psi _{\eta }(z,t) = \left({1 \over \pi }\right)^{1/2}
\exp \left\{-{1\over 2}\left(e^{-2\eta }u^{2} +
e^{2\eta }v^{2}\right)\right\} .
\end{equation}
We note here that the transition from Eq.(\ref{13}) to Eq.(\ref{14})
is a squeeze transformation.  The wave function of Eq.(\ref{13}) is
distributed within a circular region in the $u v$ plane, and thus in
the $z t$ plane.  On the other hand, the wave function of Eq.(\ref{14})
is distributed in an elliptic region.  This ellipse is a ``squeezed''
circle with the same area as the circle, as is illustrated in
Fig.~\ref{f.lcone}.

For many years, we have been interested in combining quantum mechanics
with special relativity.  One way to achieve this goal is to combine
the quantum mechanics of Fig.~\ref{f.quantum} and the relativity of
Fig.~\ref{f.lcone} to produce a covariant picture of
Fig.~\ref{f.ellipse}.  We are now ready to exploit physical consequence
of the Lorentz-squeezed quantum mechanics of Fig.~\ref{f.ellipse}.

\section{Feynman's Parton Picture}\label{parton}

It is safe to believe that hadrons are quantum bound states of quarks having
localized probability distribution.  As in all bound-state cases, this
localization condition is responsible for the existence of discrete mass
spectra.  The most convincing evidence for this bound-state picture is the
hadronic mass spectra which are observed in high-energy
laboratories~\cite{fkr71,knp86}.
However, this picture of bound states is applicable only to observers in the
Lorentz frame in which the hadron is at rest.  How would the hadrons appear
to observers in other Lorentz frames?  More specifically, can we use the
picture of Lorentz-squeezed hadrons discussed in Sec.~\ref{covham}.

Proton's radius is $10^{-5}$ of that of the hydrogen atom. Therefore,
it is not unnatural to assume that the proton has a point charge in
atomic physics.  However, while carrying out experiments on electron
scattering from proton targets, Hofstadter in 1955 observed that the
proton charge is spread out~\cite{hofsta55}.  In this experiment, an
electron emits a virtual photon, which then interacts with the proton.
If the proton consists of quarks
distributed within a finite space-time region, the virtual photon will
interact with quarks which carry fractional charges.  The scattering
amplitude will depend on the way in which quarks are distributed within
the proton.  The portion of the scattering amplitude which describes
the interaction between the virtual photon and the proton is called the
form factor.

Although there have been many attempts to explain this phenomenon
within the framework of quantum field theory, it is quite natural to
expect that the
wave function in the quark model will describe the charge distribution.  In
high-energy experiments, we are dealing with the situation in which the
momentum transfer in the scattering process is large.  Indeed, the
Lorentz-squeezed wave functions lead to the correct behavior of the hadronic
form factor for large values of the momentum transfer~\cite{fuji70}.

While the form factor is the quantity which can be extracted from the
elastic scattering, it is important to realize that in high-energy
processes, many particles are produced in the final state.  They are called
inelastic processes.  While the elastic process is described by the total
energy and momentum transfer in the center-of-mass coordinate system, there
is, in addition, the energy transfer in inelastic scattering.  Therefore, we
would expect that the scattering cross section would depend on the energy,
momentum transfer, and energy transfer.  However, one prominent feature in
inelastic scattering is that the cross section remains nearly constant for a
fixed value of the momentum-transfer/energy-transfer ratio.  This phenomenon
is called ``scaling''~\cite{bj69}.

\begin{figure}[thb] 
\centerline{\psfig{figure=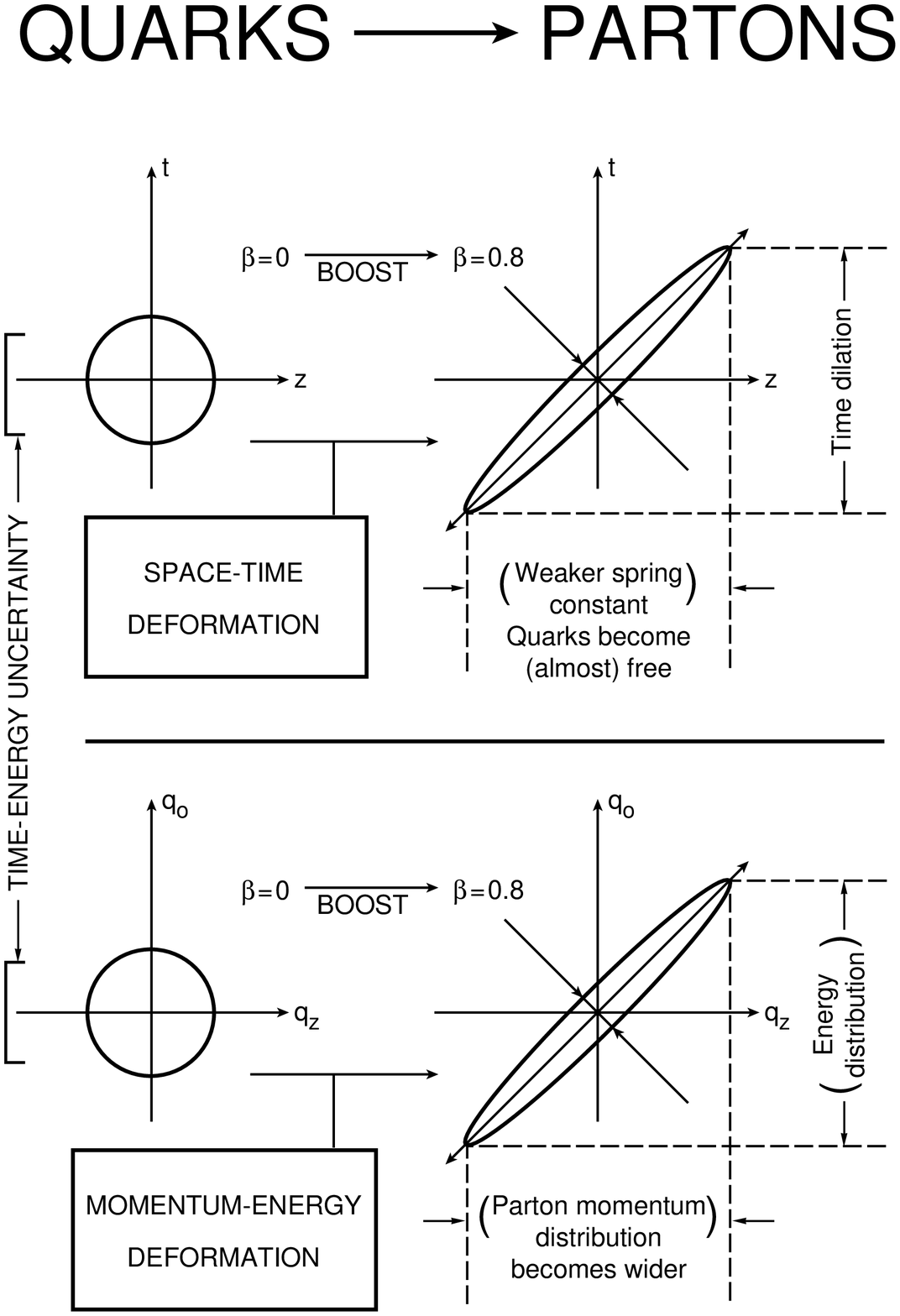,angle=0,height=140mm}}
\vspace{5mm}
\caption{Lorentz-squeezed space-time and momentum-energy wave functions.
As the hadron's speed approaches that of light, both wave functions
become concentrated along their respective positive light-cone axes.
These light-cone concentrations lead to Feynman's parton
picture.}\label{f.parton}
\end{figure}

In order to explain the scaling behavior in inelastic scattering,
Feynman in 1969 observed that a fast-moving hadron can be regarded as a
collection of many ``partons'' whose properties do not appear to be
identical to those of quarks~\cite{fey69}.  For example, the number of
quarks inside a static proton is three, while the number of partons in a
rapidly moving proton appears to be infinite.  The question then is how
the proton looking like a bound state of quarks to one observer can appear
different to an observer in a different Lorentz frame?  Feynman made the
following systematic observations.

\begin{itemize}

\item[ a).] The picture is valid only for hadrons moving with velocity
     close to that of light.

\item[ b).] The interaction time between the quarks becomes dilated,
  and partons behave as free independent particles.

\item[ c).] The momentum distribution of partons becomes widespread as
  the hadron moves very fast.

\item[ d).] The number of partons seems to be infinite or much larger
  than that of quarks.

\end{itemize}

\noindent Because the hadron is believed to be a bound state of two or
three quarks, each of the above phenomena appears as a paradox, particularly b) and
c) together.  We would like to resolve this paradox using the covariant
harmonic oscillator formalism.

For this purpose, we need a momentum-energy wave function.  If the quarks
have the four-momenta $p_{a}$ and $p_{b}$, we can construct two independent
four-momentum variables~\cite{fkr71}
\begin{equation}
P = p_{a} + p_{b} , \qquad q = \sqrt{2}(p_{a} - p_{b}) .
\end{equation}
The four-momentum $P$ is the total four-momentum and is thus the hadronic
four-momentum.  $q$ measures the four-momentum separation between the quarks.

We expect to get the momentum-energy wave function by taking the Fourier
transformation of Eq.(\ref{14}):
\begin{equation}\label{fourier}
\phi_{\eta }(q_{z},q_{0}) = \left({1 \over 2\pi }\right)
\int \psi_{\eta}(z, t) \exp{\left\{-i(q_{z}z - q_{0}t)\right\}} dx dt .
\end{equation}
Let us now define the momentum-energy variables in the light-cone coordinate
system as
\begin{equation}\label{conju}
q_{u} = (q_{0} - q_{z})/\sqrt{2} ,  \qquad
q_{v} = (q_{0} + q_{z})/\sqrt{2} .
\end{equation}
In terms of these variables, the Fourier transformation of
Eq.(\ref{fourier}) can be written as
\begin{equation}\label{fourier2}
\phi_{\eta }(q_{z},q_{0}) = \left({1 \over 2\pi }\right)
\int \psi_{\eta}(z, t) \exp{\left\{-i(q_{u} u + q_{v} v)\right\}} du dv .
\end{equation}
The resulting momentum-energy wave function is
\begin{equation}\label{phi}
\phi_{\eta }(q_{z},q_{0}) = \left({1 \over \pi }\right)^{1/2}
\exp\left\{-{1\over 2}\left(e^{-2\eta}q_{u}^{2} +
e^{2\eta}q_{v}^{2}\right)\right\} .
\end{equation}
Since we are using here the harmonic oscillator, the mathematical form
of the above momentum-energy wave function is identical to that of the
space-time wave function.  The Lorentz squeeze properties of these wave
functions are also the same, as are indicated in Fig.~\ref{f.parton}.

When the hadron is at rest with $\eta = 0$, both wave functions behave
like those for the static bound state of quarks.  As $\eta$ increases,
the wave functions become continuously squeezed until they become
concentrated along their respective positive light-cone axes.  Let us
look at the z-axis projection of the space-time wave function.  Indeed,
the width of the quark distribution increases as the hadronic speed
approaches that of the speed of light.  The position of each quark
appears widespread to the observer in the laboratory frame, and the
quarks appear like free particles.

Furthermore, interaction time of the quarks among themselves become
dilated.  Because the wave function becomes wide-spread, the distance
between one end of the harmonic oscillator well and the other end
increases as is indicated in Fig.~\ref{f.parton}.  This effect,
first noted by Feynman~\cite{fey69}, is universally observed in
high-energy hadronic experiments.  The period is oscillation is
increases like $e^{\eta}$.  On the other hand, the interaction time
with the external signal, since it is moving in the direction opposite
to the direction of the hadron, it travels along the negative
light-cone axis.  If the hadron contracts along the negative
light-cone axis, the interaction time decreases by $e^{-\eta}$.
The ratio of the interaction time to the oscillator period becomes
$e^{-2\eta}$.  The energy of each proton coming out of the Fermilab
accelerator is $900 GeV$.  This leads the ratio to $10^{-6}$.  This
is indeed a small number.  The external signal is not able to sense
the interaction of the quarks among themselves inside the hadron.

The momentum-energy wave function is just like the space-time wave
function.  The longitudinal momentum distribution becomes wide-spread
as the hadronic speed approaches the velocity of light.  This is in
contradiction with our expectation from nonrelativistic quantum
mechanics that the width of the momentum distribution is inversely
proportional to that of the position wave function.  Our expectation
is that if the quarks are free, they must have their sharply defined
momenta, not a wide-spread distribution.  This apparent
contradiction presents to us the following two fundamental questions:

\begin{itemize}

\item[a)].  If both the spatial and momentum distributions become
      widespread
      as the hadron moves, and if we insist on Heisenberg's uncertainty
      relation, is Planck's constant dependent on the hadronic velocity?

\item[b)].  Is this apparent contradiction related to another apparent
      contradiction that the number of partons is infinite while there
      are only two or three quarks inside the hadron?

\end{itemize}

The answer to the first question is ``No'', and that for the second
question is ``Yes''.  Let us answer the first question which is related
to the Lorentz invariance of Planck's constant.  If we take the product
of the width of the longitudinal momentum distribution and that of the
spatial distribution, we end up with the relation
\begin{equation}
<z^{2}><q_{z}^{2}> = (1/4)[\cosh(2\eta)]^{2}  .
\end{equation}
The right-hand side increases as the velocity parameter increases.
This could lead us to an erroneous conclusion that Planck's constant
becomes dependent on velocity.  This is not correct, because the
longitudinal momentum variable $q_{z}$ is no longer conjugate to the
longitudinal position variable when the hadron moves.

In order to maintain the Lorentz-invariance of the uncertainty product,
we have to work with a conjugate pair of variables whose product does
not depend on the velocity parameter.  Let us go back to Eq.(\ref{conju})
and Eq.(\ref{fourier2}).  It is quite clear that the light-cone variable
$u$ and $v$ are conjugate to $q_{u}$ and $q_{v}$ respectively.  It is
also clear that the distribution along the $q_{u}$ axis shrinks as the
$u$-axis distribution expands.  The exact calculation leads to
\begin{equation}
<u^{2}><q_{u}^{2}> = 1/4 , \qquad  <v^{2}><q_{v}^{2}> = 1/4  .
\end{equation}
Planck's constant is indeed Lorentz-invariant.

Let us next resolve the puzzle of why the number of partons appears to
be infinite while there are only a finite number of quarks inside the
hadron.  As the hadronic speed approaches the speed of light, both the
x and q distributions become concentrated along the positive light-cone
axis.  This means that the quarks also move with velocity very close
to that of light.  Quarks in this case behave like massless particles.

\begin{figure}[thb]
\centerline{\psfig{figure=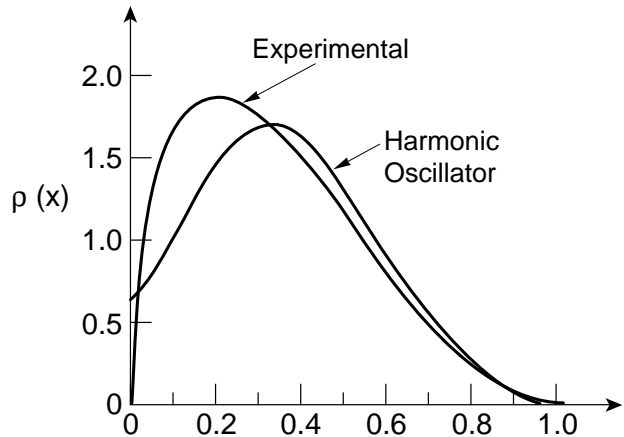,angle=0,height=60mm}}
\caption{Parton distribution.  It is possible to calculate the parton
distribution from the Lorentz-boosted oscillator wave function.
This theoretical curve is compared with the experimental
curve.}\label{f.hussar}
\end{figure}


We then know from statistical mechanics that the number of massless
particles is not a conserved quantity.  For instance, in black-body
radiation, free light-like particles have a widespread momentum
distribution.  However, this does not contradict the known principles
of quantum mechanics, because the massless photons can be divided into
infinitely many massless particles with a continuous momentum
distribution.

Likewise, in the parton picture, massless free quarks have a wide-spread
momentum distribution.  They can appear as a distribution of an
infinite number of free particles.  These free massless particles are the
partons.  It is possible to measure this distribution in high-energy
laboratories, and it is also possible to calculate it using the covariant
harmonic oscillator formalism.  We are thus forced to compare these two
results.  Indeed, according to Hussar's calculation~\cite{hussar81},
the Lorentz-boosted oscillator wave function produces a reasonably
accurate parton distribution, as indicated in Fig.~\ref{f.hussar}

\section*{Concluding Remarks}
In this report, we have considered a string consisting only of two
particles bounded together by an oscillator potential.  The essence
of the problem was to construct a quantum mechanics of harmonic
oscillators which can be Lorentz-transformed.
We achieved this purpose by remodeling the oscillator formalism of
Feynman, Kislinger and Ravndal.  Their Lorentz-invariant equation
has a covariant set of solutions which is consistent with the
existing principles of quantum mechanics and special relativity.

From these wave wave functions, it is possible to construct a
representation of Wigner's $O(3)$-like little group governing the
internal space-time symmetries of relativistic particles with
non-zero mass.  In order to illustrate the difference between the
little group for massive particles from that for massless particles,
we have given a comprehensive review of the little groups for massive
and massless particles.  We have discussed also the contraction
procedure in which the $E(2)$-like little group for massless particles
is obtained from the $O(3)$-like little group for massive particles.
We have given a comprehensive review of the contents of Table I.

Let us go back to the issue of strings.  As we noted earlier in this
paper, the string is a limiting case of discrete sets of mass points.
We can consider two limiting cases, namely the continuous string and
two-particle string.  There also is a possibility of strings of
discrete sets of particles, or ``polymers of point-like
constituents''\cite{thorn97}.  These different strings might take
different mathematical forms, but they should all share the space-time
symmetry.  Thus, the quickest way to study this symmetry is to use
the simplest mathematical technique which the two-pearl string
provides.

\section*{Acknowledgments}
The author would like to thank Academician A. A. Logunov and the members
of the organizing committee for inviting him to visit the Institute
of High Energy Physics at Protvino and participate in the 22nd
International Workshop on the Fundamental Problems of High Energy
Physics.  The original title of this paper was ``Two-bead Strings,''
but Professor V. A. Petrov changed it to ``Two-pearl Strings'' when he
was introducing the author and the title to the audience.  He was the
chairman of the session in which the author presented this paper.

\begin{appendix}

\section{Contraction of O(3) to E(2)}\label{o3e2}
In this Appendix, we explain what the $E(2)$ group is.  We then
explain how we can obtain this group from the three-dimensional
rotation group by making a flat-surface or cylindrical approximation.
This contraction procedure will give a clue to obtaining the $E(2)$-like
symmetry for massless particles from the $O(3)$-like symmetry for
massive particles by making the infinite-momentum limit.

The $E(2)$ transformations consist of rotation and two translations on
a flat plane.  Let us start with the  rotation matrix applicable to
the column vector $(x, y, 1)$:
\begin{equation}\label{rot}
R(\theta) = \pmatrix{\cos\theta & -\sin\theta & 0 \cr
\sin\theta & \cos\theta & 0 \cr 0 & 0 & 1} .
\end{equation}
Let us then consider the translation matrix:
\begin{equation}
T(a, b) = \pmatrix{1 & 0 & a \cr 0 & 1 & b \cr 0 & 0 & 1} .
\end{equation}
If we take the product $T(a, b) R(\theta)$,
\begin{equation}\label{eucl}
E(a, b, \theta) = T(a, b) R(\theta) =
\pmatrix{\cos\theta & -\sin\theta & a \cr
\sin\theta & \cos\theta & b \cr 0 & 0 & 1} .
\end{equation}
This is the Euclidean transformation matrix applicable to the
two-dimensional $x y$ plane.  The matrices $R(\theta)$ and $T(a,b)$
represent the rotation and translation subgroups respectively.  The
above expression is not a direct product because $R(\theta)$ does not
commute with $T(a,b)$.  The translations constitute an Abelian
invariant subgroup because two different $T$ matrices commute with
each other, and because
\begin{equation}
R(\theta) T(a,b) R^{-1}(\theta) = T(a',b') .
\end{equation}
The rotation subgroup is not invariant because the conjugation
$$T(a,b) R(\theta) T^{-1}(a,b)$$
does not lead to another rotation.

We can write the above transformation matrix in terms of generators.
The rotation is generated by
\begin{equation}
J_{3} = \pmatrix{0 & -i & 0 \cr i & 0 & 0 \cr 0 & 0 & 0} .
\end{equation}
The translations are generated by
\begin{equation}
P_{1} = \pmatrix{0 & 0 & i \cr 0 & 0 & 0 \cr 0 & 0 & 0} , \qquad
P_{2} = \pmatrix{0 & 0 & 0 \cr 0 & 0 & i \cr 0 & 0 & 0} .
\end{equation}
These generators satisfy the commutation relations:
\begin{equation}\label{e2com}
[P_{1}, P_{2}] = 0 , \qquad [J_{3}, P_{1}] = iP_{2} ,
\qquad [J_{3}, P_{2}] = -iP_{1} .
\end{equation}
This $E(2)$ group is not only convenient for illustrating the groups
containing an Abelian invariant subgroup, but also occupies an
important place in constructing representations for the little
group for massless particles, since the little group for massless
particles is locally isomorphic to the above $E(2)$ group.

The contraction of $O(3)$ to $E(2)$ is well known and is often called
the Inonu-Wigner contraction~\cite{inonu53}.  The question is whether
the $E(2)$-like little group can be obtained from the $O(3)$-like
little group.  In order to answer this question, let us closely look
at the original form of the Inonu-Wigner contraction.  We start with
the generators of $O(3)$.  The $J_{3}$ matrix is given in Eq.(\ref{j3}),
and
\begin{equation}\label{o3gen}
J_{2} = \pmatrix{0&0&i\cr0&0&0\cr-i&0&0} , \qquad
J_{3} = \pmatrix{0&-i&0\cr i &0&0\cr0&0&0} .
\end{equation}
The Euclidean group $E(2)$ is generated by $J_{3}, P_{1}$ and $P_{2}$,
and their Lie algebra has been discussed in Sec.~\ref{intro}.

Let us transpose the Lie algebra of the $E(2)$ group.  Then $P_{1}$ and
$P_{2}$ become $Q_{1}$ and $Q_{2}$ respectively, where
\begin{equation}
Q_{1} = \pmatrix{0&0&0\cr0&0&0\cr i &0&0} , \qquad
Q_{2} = \pmatrix{0&0&0\cr0&0&0\cr0&i&0} .
\end{equation}
Together with $J_{3}$, these generators satisfy the
same set of commutation relations as that for
$J_{3}, P_{1}$, and $P_{2}$ given in Eq.(\ref{e2com}):
\begin{equation}
[Q_{1}, Q_{2}] = 0 , \qquad [J_{3}, Q_{1}] = iQ_{2} , \qquad
[J_{3}, Q_{2}] = -iQ_{1} .
\end{equation}
These matrices generate transformations of a point on a circular
cylinder.  Rotations around the cylindrical axis are generated by
$J_{3}$.  The matrices $Q_{1}$ and $Q_{2}$ generate translations along
the direction of $z$ axis.  The group generated by these three matrices
is called the {\it cylindrical group}~\cite{kiwi87jm,kiwi90jm}.

We can achieve the contractions to the Euclidean and cylindrical groups
by taking the large-radius limits of
\begin{equation}\label{inonucont}
P_{1} = {1\over R} B^{-1} J_{2} B ,
\qquad P_{2} = -{1\over R} B^{-1} J_{1} B ,
\end{equation}
and
\begin{equation}
Q_{1} = -{1\over R}B J_{2}B^{-1} , \qquad
Q_{2} = {1\over R} B J_{1} B^{-1} ,
\end{equation}
where
\begin{equation}\label{bmatrix}
B(R) = \pmatrix{1&0&0\cr0&1&0\cr0&0&R}  .
\end{equation}
The vector spaces to which the above generators are applicable are
$(x, y, z/R)$ and $(x, y, Rz)$ for the Euclidean and cylindrical groups
respectively.  They can be regarded as the north-pole and equatorial-belt
approximations of the spherical surface respectively~\cite{kiwi87jm}.

\section{Contraction of O(3)-like to E(2)-like Little Groups}\label{contrac}

Since $P_{1} (P_{2})$ commutes with $Q_{2} (Q_{1})$, we can consider the
following combination of generators.
\begin{equation}
F_{1} = P_{1} + Q_{1} , \qquad F_{2} = P_{2} + Q_{2} .
\end{equation}
Then these operators also satisfy the commutation relations:
\begin{equation}\label{commuf}
[F_{1}, F_{2}] = 0 , \qquad [J_{3}, F_{1}] = iF_{2} , \qquad
[J_{3}, F_{2}] = -iF_{1} .
\end{equation}
However, we cannot make this addition using the three-by-three matrices
for $P_{i}$ and $Q_{i}$ to construct three-by-three matrices for $F_{1}$
and $F_{2}$, because the vector spaces are different for the $P_{i}$ and
$Q_{i}$ representations.  We can accommodate this difference by creating
two different $z$ coordinates, one with a contracted $z$ and the other
with an expanded $z$, namely $(x, y, Rz, z/R)$.  Then the generators
become
\begin{equation}
P_{1} = \pmatrix{0&0&0&i\cr0&0&0&0\cr0&0&0&0\cr0&0&0&0} , \qquad
P_{2} = \pmatrix{0&0&0&0\cr0&0&0&i\cr0&0&0&0\cr0&0&0&0} .
\end{equation}
\begin{equation}
Q_{1} = \pmatrix{0&0&0&0\cr0&0&0&0\cr i &0&0&0\cr0&0&0&0} , \qquad
Q_{2} = \pmatrix{0&0&0&0\cr0&0&0&0\cr0&i&0&0\cr0&0&0&0} .
\end{equation}
Then $F_{1}$ and $F_{2}$ will take the form
\begin{equation}\label{f1f2}
F_{1} = \pmatrix{0&0&0&i\cr0&0&0&0\cr i &0&0&0\cr0&0&0&0} , \qquad
F_{2} = \pmatrix{0&0&0&0\cr0&0&0&i\cr0&i&0&0\cr0&0&0&0} .
\end{equation}
The rotation generator $J_{3}$ takes the form of Eq.(\ref{j3}).
These four-by-four matrices satisfy the E(2)-like commutation relations
of Eq.(\ref{commuf}).

Now the $B$ matrix of Eq.(\ref{bmatrix}), can be expanded to
\begin{equation}\label{bmatrix2}
B(R) = \pmatrix{1&0&0&0\cr0&1&0&0\cr0&0&R&0\cr0&0&0&1/R} .
\end{equation}
If we make a similarity transformation on the above form using the matrix
\begin{equation}\label{simil}
\pmatrix{1&0&0&0\cr0&1&0&0\cr0&0&1/\sqrt{2} &-1/\sqrt{2}
\cr0&0&1/\sqrt{2}&1/\sqrt{2}} ,
\end{equation}
which performs a 45-degree rotation of the third and fourth coordinates,
then this matrix becomes
\begin{equation}\label{simil2}
\pmatrix{1&0&0&0\cr0&1&0&0\cr0&0 & \cosh\eta & \sinh\eta
\cr0 & 0 & \sinh\eta & \cosh\eta} ,
\end{equation}
with $R = e^\eta$.  This form is the Lorentz boost matrix along the $z$
direction.  If we start with the set of expanded rotation generators
$J_{3}$ of Eq.(\ref{j3}), and
perform the same operation as the original Inonu-Wigner contraction
given in Eq.(\ref{inonucont}), the result is
\begin{equation}
N_{1} = {1\over R} B^{-1} J_{2} B ,
\qquad N_{2} = -{1\over R} B^{-1} J_{1} B ,
\end{equation}
where $N_{1}$ and $N_{2}$ are given in Eq.(\ref{n1n2}).  The generators
$N_{1}$ and $N_{2}$ are the contracted $J_{2}$ and $J_{1}$ respectively
in the infinite-momentum/zero-mass limit.

\end{appendix}

\end{document}